# Assessing the Impact of the Physical Environment on Comfort and Job Satisfaction in Offices

Francisco Reyne-Pugh, José Pulgar, Alex Godoy-Faúndez, Mario Alvarado-Rybak, Cristóbal Galbán-Malagón. (Dated: January 13, 2020)

## ABSTRACT

*This paper develops a model that allows to analyze the physical parameters that determine the degree of environmental comfort of employees in offices. Parameters such as air quality, noise, thermal environment, and lighting are considered. This model was developed through the use of partial least squares structural equation models (PLS-SEM). Formative indicators (which cause the construct) and reflective indicators (caused or affected by the construct) were used, following the methodology proposed by Hair et al. (2014). The model was estimated using data obtained in surveys conducted in aeronautical control offices in Chile (DASA-DGAC), The model allows to evaluate the influence that environmental comfort has on people's job satisfaction. The results indicate that the environmental parameters used significantly influence environmental comfort, explaining 70.2% of its variance. In addition, it was obtained that the influence of noise on environmental comfort proved to be greater than that of the rest of the environmental parameters studied, followed by air quality. On the other hand, it was empirically proven that environmental comfort has a significant influence on job satisfaction, where the environmental parameters used explain the variance of job satisfaction by 28.9%.*

## INTRODUCTION

Today the population living in urban centers spend more than 90% of the time in indoor spaces (Rey & Velasco, 2007). Such is the case of cities, where 4,027 million people live, about half of the world's population (World Bank, 2017). In this context the understanding of how are structured the perceptions of indoor environment, is important because as indicated by Clausen and Wyon (2008), environmental comfort has an influence on the physical and psychological well-being of the people. By other hand, if the welfare of workers is affected then productivity, competitiveness and sustainability of the companies are affected (OMS, 2013). That makes an opportunity to study the offices, as an indoor working space. The indoor environmental comfort is determined by environmental physical conditions such as lighting, thermal environment, noise (Li Huang & Qin Ouyang, 2012; Horie. Sakurai, Narguchi, & Matsubara, 1985) and air quality (Clausen , 1994; Rohles, 1989; Woods, 1987; Wong, Mui, & Hui, 2008).

Several studies show how physical indicators influence people, measuring levels of environmental comfort through surveys. Which is fine because it's subjective (INP, 2012), but there are some problems related to measuring unobservable or latent variables (also called constructs) such as environmental comfort, creativity, job satisfaction, or others concepts measured through questionnaires. One problem is the use of dichotomous scales, which is a serious problem (Pett, Lackey, & Sullivan, 2003), and is used in studies of environmental comfort as the case of Wong, Mui and Hui (2008). Other problem is the lack of a central value in the measurements (Pett, Lackey, & Sullivan, 2003) as the scale used by Li Huang and Qin Ouyang (2012). Last but not least, the problem associated with not assessing the validity and reliability of the measures (Hair, Hult, Ringle, & Sarstedt, 2014; Chiang, Martín, & Núñez, 2010; APA, AERA, & NCME, 1999), for example the case of Wong, Miu and Hui (2008), Bernardi and Kowaltowski (2006), Li Huang and Qin Ouyang (2012), Clausen (1994) and Rohles (1989).

The aim of this study was to propose a model to measure environmental comfort through the use of Partial Least Squares Structural Equation Modeling (PLS-SEM), which allows to evaluate linear relationships between latent variables, and to analyze the validity and reliability of results. We choose PLS-SEM instead of Covariance Based Structural Equation Modeling (CB-SEM) because PLS-SEM has small sample size and non-normal data distribution requirements and has the ability to include formative indicators (Diamantopoulos &Riefler, 2011; Diamantopoulos, Riefler, & Roth, 2008; Diamantopoulos &Winklhofer, 2001; Gudergan, Ringle, Wende, & Will, 2008; Jarvis, MacKenzie, & Podsakoff, 2003; MacKenzie, Podsakoff, & Jarvis, 2005; MacKenzie, Podsakoff, & Podsakoff , 2011).The hypotheses of our research were: first, that there was a positive and significant relationship between environmental comfort and job satisfaction, and second, that there were environmental indicators that influence environmental comfort more than others.

# METHODS

We generate an Index of environmental comfort that was constructed and validated following the recommendations given by Hair, Ringle and Sarstedt (2014).

**Data Characteristics:** Cross-sectional observational study, using PLS-SEM in SmartPLS (Ringle, Wende, & Will, 2005). Since environmental comfort is subjective (INP, 2012), it was measured by surveys. These questionnaires were applied to employees of the General Direction of Aeronautics of Chile who worked in offices and who decided to do it voluntarily. The participants were 45 volunteers. Only one participant answered the survey incompletely and was discarded from the study as part of the treatment of lost data, in accordance to avoid problems of results bias (Hair, Ringle, &Sarstedt, 2014). For the validation of the instrument, the number of individuals was adapted to what was recommended by Kline (2005) and Chin (1998) cited by Christophersen and Konradt (2006) and it is more than ten times the number of formative indicators. We had 44 cases with four formative indicators.

Mean, standard deviation, minimum and maximum values were calculated for each indicator.

**Construct definition:** As Olesen and Seelen (1993) point out referring to the investigation of environmental comfort: "It is, however, important to realize that all factors are related and comfort perceived by occupants is a combination of all the factors". We define environmental comfort as "the mental condition in which satisfaction is expressed with respect to the physical conditions of the surroundings or environment, specifically those environmental physical conditions associated with noise, lighting, air quality, and thermal environment". We generate this definition inspired by the definition of thermal comfort mentioned in the norm UNE-EN ISO 7730 (2006) which says that thermal comfort is: "a mental condition in which the satisfaction with the thermal environment is expressed".

**Indicators definition:** From our environmental comfort definition, we generated a single item for each of the four environmental physical conditions (noise, lighting, air quality, and thermal environment) as a formative indicator, using a Likert scale (Pett, Lackey, & Sullivan, 2003) specifically a conceptual qualification scale with seven response options. The scales used were from totally dissatisfied to totally satisfied. We used two reflective indicators, one was "the degree of satisfaction with physical conditions" and the other was "the degree of satisfaction with physical environment", following the recommendations of Hair, Hult, Ringle, &Sarstedt (2014) these two reflective measures were similar aspects of the construct "environmental comfort" and summarize the content of it. To measure job satisfaction, the Overall Job Satisfaction Scale by Warr, Cook and Wall (1979) adapted to Spanish by Perez and Fidalgo (1994) was used. This scale was established as a reflective measurement of job satisfaction given the formula for its calculation (direct sum of each item, or equivalent weights). To make the measurements compatible in both scales (environmental comfort and job satisfaction) all the items were of seven response options, as Pett, Lackey and Sullivan (2003) recommended.

# Model Validity - Convergent validity & Discriminant validity

**Convergent validity: Redundancy analysis of the environmental comfort model**

As formative measurement models should be evaluated cautiously, we followed the recommendations given by Hair, Hult, Ringle and Sarstedt (2014) in the context of PLS-SEM. Environmental comfort was linked to itself between a Formative model and a Reflective model. See Figure 1.

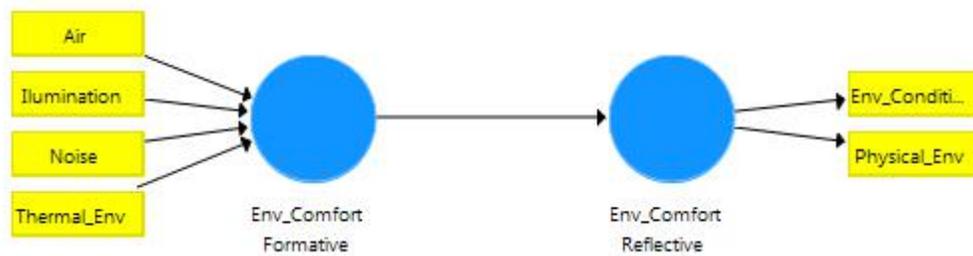

*Figure 1: Redundancy Analysis*

With the information of the indicators of this model we analyzed data distribution, association between formative and reflective indicators, collinearity of formative indicators, reliability of reflective indicators of environmental comfort, and convergent validity of the environmental comfort reflective model, and convergent validity.

In the PLS-SEM Algorithm we used the following settings: a path weighting scheme, a maximum iteration of 300, and a stop criterion of $10^{-7}$. Path weighting scheme provides the highest R² value for endogenous latent variables and is generally applicable for all kinds of PLS path model specifications and estimations (Ringle, Wende, & Becker, 2015).

In the Bootstrapping procedure we used the following settings: 5000 subsamples, Complete bootstrapping results, Bias-corrected and Accelerated (BCa) Bootstrap, Two Tailed test type, and significance level of 0,05. We follow the advice of Ringle, Wende, & Becker (2015) to use "Bias-Corrected and Accelerated (BCa) Bootstrap" as it is the most stable method that does not need excessive computing time, a large number of bootstrap subsamples (e.g., 5,000) for final results, and parallel processing to reduce computation time.

**Discriminant validity: evaluating the relationship of Environmental Comfort with Job Satisfaction**

The redundancy model shown in the Figure 1 was linked to the measurement model of job satisfaction. See Figure 2.

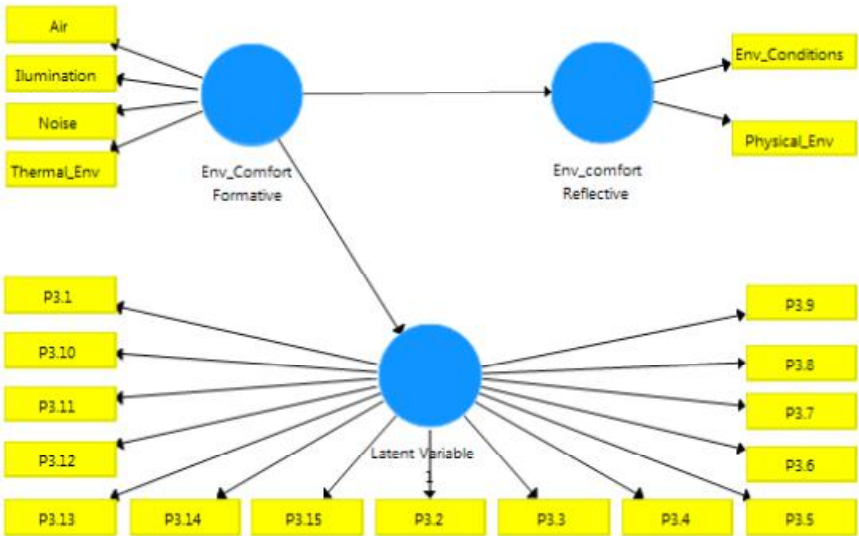

*Figure 2: Discriminant Validity Model*

This correspond to our discriminant validity model, in this model we analyzed the Reliability of reflective indicators of job satisfaction, Convergent validity of the Job Satisfaction reflective model, and finally the Discriminant validity of the environmental comfort model.

We used the same settings in the PLS-SEM algorithm and in the Bootstrapping procedure as we did for the Redundancy analysis.

## Index & Scale of Environmental Comfort in Offices

**Construction of the Environmental Comfort Index for offices:** The construction of the index was done based on the outer weights of the formative indicators obtained in the redundancy analysis (Diamantopoulos & Winklhofer, 2001). The value of the Environmental Comfort Index (ECI) was developed using formula 1. See Equation below.

$$Environmental\ Comfort\ Index\ (ECI) = \sum(FI_i \cdot OW_i)$$  **Formula 1.**

Where: FIi: value of the Formative Indicator i, OWi: Outer weight of the indicator i.

**Construction of the Environmental Comfort Scale for offices**: The reflective measure of environmental comfort was equivalent to the average of the two reflective indicators, as shown in Formula 2 shown below:

$$Environmental\ Comfort\ Scale\ (ECS) = \sum(IR_i)/m \qquad \textbf{Formula 2.}$$

Where: IRi: value of the reflective indicator i, m: number of reflective indicators.

# RESULTS

The obtained age range was from 21 to 76 years with an average of 44.4 years. Regarding the time working in the institution the range was from 1 month to 48 years, with an average of 19.5 years. The obtained average and standard deviation of each environmental indicator are shown in Figure 3.

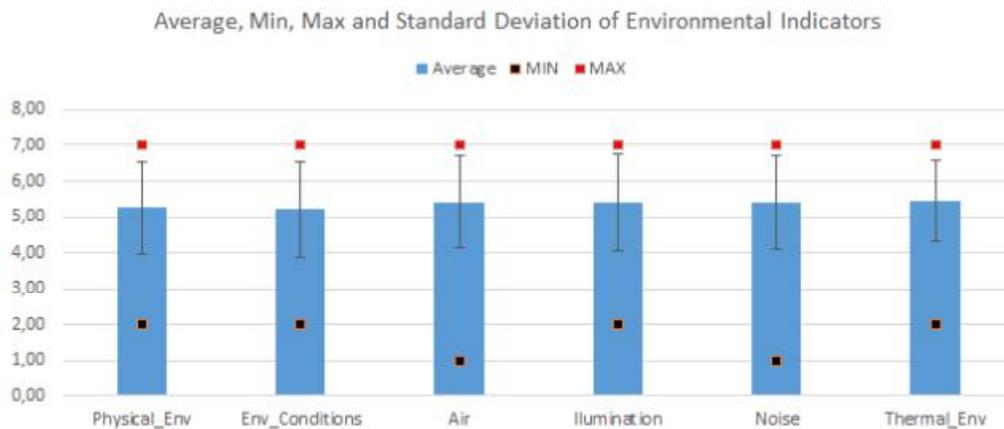

*Figure 3: Average and Standard deviation of Environmental Comfort Indicators*

The results obtained in the PLS-SEM and Bootstrapping algorithms are shown as follows: the convergent validity results of the redundancy analysis is shown in Table A, the convergent validity results of the job satisfaction and environmental comfort model are shown in Table B, the discriminant validity results of the job satisfaction and environmental comfort model is shown in Table C, Table D and Table E.

Table A.

| Convergent validity: Redundancy analysis | | | |
|---|---|---|---|
| Analysis | Statistic | Acceptance values | Result |
| Data Distribution | Shapiro-Wilk statistic (n<50) (Mohd & Bee, 2011) | Normality: p-value > 0,05 | All p-values were lower than 0.05 |
| Association between indicators | if Normal, we calculate the Pearson correlation, if not Spearman correlation. | significant relationship: p-values < 0,05 | All relationship between formative and reflective indicators were significant |
| Collinearity of Formative indicators | Variance inflation factor (VIF) | VIF values less than 5. | All formative indicator had VIF < 5 (Air: 2.787, Ilumination: 1.627, Noise: 1.923, Thermal environment: 1.340) |
| Reliability of reflective indicators | Cronbach's alpha, composite reliability, and the outer loading values | All this values should be greater than 0.708. | Composite realibility value was 0,874. Conbrach alpha was 0.712, and all outer loadings were 0.895 and 0.867 |
| Convergent validity: Reflective Model | Average Variance Extracted (AVE) | Greater than 0.50. | 0.776 |
| Latent variable relationship | R Square, f Square, rho_A | Greater than 0.64 | 0.702 (p-value: 0.000), 2.477 (p-value: 0.039), 0.713 (p-value: 0.000) |

Table B.

| Environmental comfort & Job Satisfaction | | | |
|---|---|---|---|
| Reliability of reflective indicators of job satisfaction | Composite reliability and Cronbach alpha | Both values greater than 0.708 | Composite reliability of 0,920 and a Cronbach's Alpha of 0,910 |
| Convergent validity of the Job Satisfaction reflective model | The value of the Average Variance Extracted (AVE) for job satisfaction | Must be above the critical value of 0.50 | At first the AVE obtained for job satisfaction was 0.454. The elimination of four indicators was necessary (3.9, 3.7, 3.15 and 3.13), achieving an AVE value of 0.516. |
| Latent variable relationship | R Square, f Square, rho_A | all values | 0.289 (p-value: 0.002), 0.407 (p-value:0.227), 0.977 (p-value: 0.000) |

Table C.

| Discriminant validity of the environmental comfort model | Fornell Larcker criterion | Diagonal values have to be greater than the others in the same column and row | See Fornell Larcker Criterion table |
|---|---|---|---|
| | Crossloadings | Cross Loading should be less than the loading on the main construct | See Cross Loadings table |
| | Heterotrait Monotrait Ratio (HTMT) | HTMT value below 0.90 | 0.551 |

Table D.

**TABLE: Fornell Larcker criterion**

|  | Env_Comfort Formative Model | Env_Comfort Reflective Model | Job Satisfaction |
|---|---|---|---|
| Env_Comfort Formative Model |  |  |  |
| Env_Comfort Reflective Model | 0.838 | 0.881 |  |
| Job Satisfaction | 0.538 | 0.541 | 0.718 |

Table E.

**TABLE: Cross Loadings**

|  | Env_Comfort Formative Model | Env_Comfort Reflective Model | Job Satisfaction |
|---|---|---|---|
| Air | *0.878* | 0.731 | 0.479 |
| Env_Conditions | 0.752 | *0.886* | 0.356 |
| Ilumination | *0.618* | 0.458 | 0.425 |
| Noise | *0.917* | 0.801 | 0.442 |
| P3.1 | 0.661 | 0.742 | *0.745* |
| P3.10 | 0.151 | 0.143 | *0.596* |
| P3.11 | 0.337 | 0.345 | *0.700* |
| P3.12 | 0.389 | 0.323 | *0.848* |
| P3.14 | 0.159 | 0.068 | *0.664* |
| P3.2 | 0.585 | 0.490 | *0.852* |
| P3.3 | 0.233 | 0.386 | *0.525* |
| P3.4 | 0.217 | 0.303 | *0.727* |
| P3.5 | 0.294 | 0.306 | *0.684* |
| P3.6 | 0.163 | 0.200 | *0.627* |
| P3.8 | 0.272 | 0.211 | *0.850* |
| Physical_Env | 0.724 | *0.876* | 0.601 |
| Thermal_Env | *0.619* | 0.477 | 0.398 |

## Construction of the Environmental Comfort Index for Offices

We standardize the outer weights of formative indicators in such a way that their sum completes the unit, dividing the value of each outer weight by their total sum. In Figure 4 the graph shows the contribution that each indicator has on Environmental comfort (outer weights).

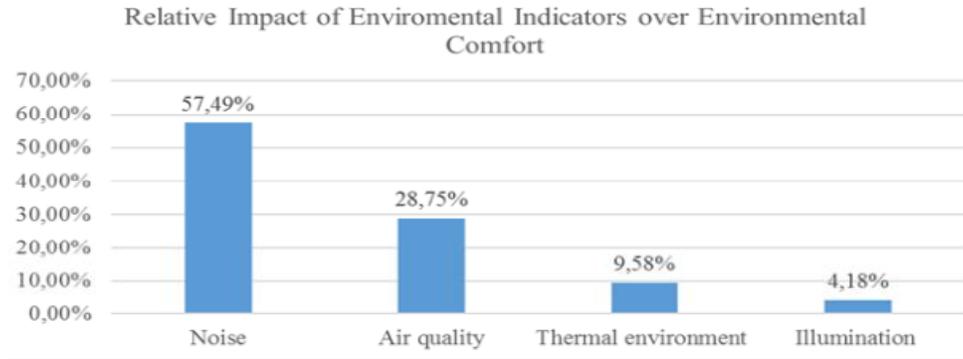

*Figure 4: Relative Impact of Environmental Indicators over Environmental Comfort*

With the outer weights showed in figure 4, the data required to complete the Formula 1, we developed the ECI.

**Other analysis**

**Comparison of Environmental Comfort Index with Environmental Comfort Scale**

The results of the values obtained using the formulas for Environmental Comfort Index and Environmental Comfort Scale for each person were illustrated in a scatter plot (Figure 5) to see their association, as well as bar graphs (Figure 6) to observe their differences. This information allows us to visualize easy linearity behavior between the two models of measurement (reflective and formative).

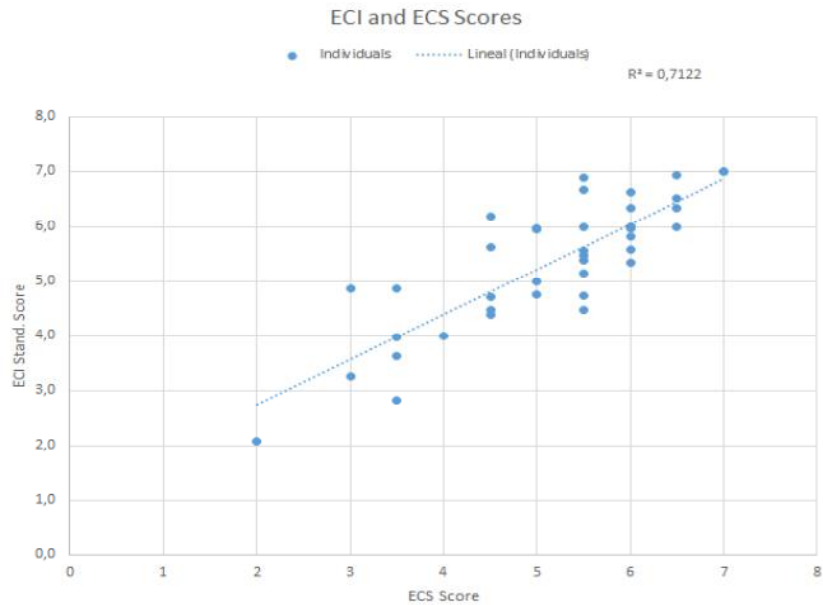

*Figure 5: ECI and ECS Scores*

As can be seen in Figure 5, there was a great association between the scores of the ECI and the ECS. The noise, the air quality, the thermal environment, and the lighting explain 71.2% of the variation level of environmental comfort of the people in this case. See Figure 6 to see absolute differences between ECI and ECS results.

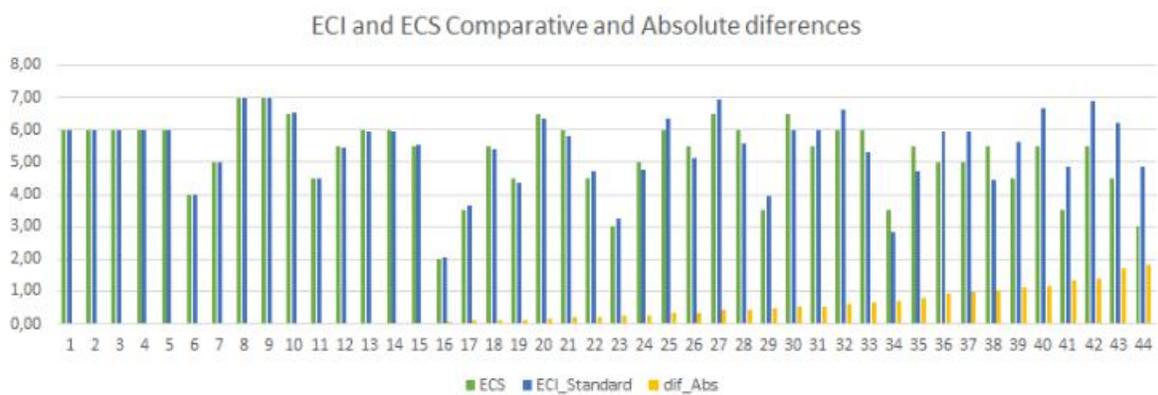

*Figure 6: ECI and ECS comparative and Absolute differences*

# DISCUSSION

We have conceptualized the construct of environmental comfort, we have measured it in a formative way to generate an index of environmental comfort and in a reflective way to generate a scale of environmental comfort. We must always consider that the criteria for the construction of indices are different from those of the construction of scales (Diamantopoulos & Winklhofer, 2001). We have addressed each of these criteria starting with the definition of environmental comfort, through the validity of the model, to the analysis of results. The collinearity analysis was favorable, the redundancy analysis confirmed the convergent validity with the non-elimination of environmental indicators.

The four environmental indicators explained significantly and to a large extent Environmental comfort ($R^2$=71,2%; r=0,84; p-value=0,001). On the other hand, the discriminant validity was verified, verifying that the environmental indicators used in the model have a significant and important relationship with job satisfaction ($R^2$=20,2%; p-value=0,001). Considering the well-known tendency of PLS-SEM to slightly underestimate structural model relationships (Dijkstra, 1983), we confirmed that environmental comfort has a big causal effect on job satisfaction.

The formative indicators proved to have very different weights on environmental comfort. This draws attention considering that the averages obtained for indicators were similar, as can be seen on Figure 3. Noise and air quality were the ones that most affect individuals, while lighting is the one that affects to a lesser extent. The difference is huge, noise affected 13,75 times more than lighting (see Formula 3). Although not all environmental indicators had a good level of significance, their presence in the model was strictly necessary because of the values of their outer loads and the existing theoretical support (Hair, 2014). Bakan (1966) mentioned that to obtain significance it is necessary to carry out studies with a greater number of individuals, because as said by Sullivan and Feinn (2012 ) with a sufficiently large sample, a statistical test will almost always demonstrate statistical significance. The relationship between p-values and strength of statistical evidence are discussed more deeply by Royall (1986), who says that sample size must also be considered when interpreting significance levels.

The results of our models have shown validity in our measurements; however, the results are not generalizable beyond the same offices and time when the data was obtained. Knowing that environmental comfort is subjective (INP, 2012) makes us believe that the outer weights of formative indicators could vary when comparing different offices and periods of time.

We consider that the instrument generated has a high potential for application in the business context, first for its simplicity of use and for the value of the information delivered, which can be used to control environmental conditions to have positive impact over job satisfaction, and second because of the increase in efficiency in the administration of resources since these will be focused on generating positive changes in those aspects that are more influential in environmental comfort and job satisfaction.

With six questions we can determine how different environmental indicators affect the global perception of Environmental comfort. This could be a solution to solve the problem discussed by Clausen (1994) about not being able to take social consensus on which improvements on environmental indicators should be prioritized when there was a limited budget.

We believe that people don't know very well how their perceptions are combined or structured in their brains, nor to what extent, which would make them unable to achieve these social consensuses. If we create a system, for example of information technologies, that incorporates our proposed model of measurement and that is able to capture information from employees constantly, it would make go from a personal choice to a social choice to make changes in the indoor environmental quality (adjusting the environment conditions), maximizing the environmental comfort of people to increase their level of job satisfaction and productivity.

We the people, each one of us are systems with different sensors that give us the information about our near environment. This information goes from our sensors to our brains, and there we compute this information and create an output in response. This output could make a physical response or/and psychological response. With our measurement we tried to take this psychological response information asking for the satisfaction level of different aspects of our environment, in the offices. We found that people compute this information in their brains in a very similar way.

Bringing workers to higher levels of satisfaction and productivity can have a negative net impact on the environment. First, considering the additional consumption of resources (energy and materials) for keeping people in ideal conditions (for adapting environmental conditions). Second, more productive people can make an industry with negative impact accelerate, increasing the negative impact generated in the same period of time. In this sense, environmental comfort would be presented as a more social than environmental alternative, however each case must be analyzed, since for example in some industries some processes could be done faster by consuming less resources for each task completed.

What happens if, for example, all companies improve their environmental performance through environmental comfort, doing the tasks more efficiently, in less time and therefore consuming less resources per result obtained. If all the companies do the same, it is possible that the effect obtained is not the desired one since a greater environmental problem could be generated globally. How? accelerating industries and markets that ultimately generate a greater negative impact on the environment. These questions generate new questions to solve, such as: environmentally speaking, to what extent should we make companies more efficient? Is there a relationship between making more efficient and making production faster? Within the context of sustainable development, we have doubts about the extent to which we must keep people in better conditions, knowing that doing so generates a series of additional resource consumptions and could make industries with negative impact more fast. Maybe answering these types of questions will help us determine how many people we should be to use the resources available in the environment to provide us with the quality of life we deserve.

We invite the scientific education community to conduct studies of environmental comfort with the proposed model applied to university or school classrooms, for their possible influence on performance and satisfaction witheducational environments.

# REFERENCES

APA, AERA, & NCME. (1999). *American Educational Research Association [AERA], American Psychological Association [APA], & National Council on Measurement in Education [NCME]. Standards for Educational and Pschylogical Testing.*

Bakan, D. (1966). The test of significance in psychological research. *Psychological Bulletin*, 66(6), 423–437. doi:10.1037/h0020412

Bernardi, N., &Kowaltowski, D. (2006). *Environmental comfort in school buildings: awareness and participation of users.* Campinas, Brasil: State University of Campinas,UNICAMP, School of Civil Engineering, Dep. of Architecture and Construction.

Chiang Vega, M., Martín Rodrigo, M. J., & Núñez Partido, A. (2010). *Relaciones entre el clima organizacional y la satisfacción laboral.* Universidad Pontíficia Comillas.

Christophersen, Timo, &Konradt, Udo. (2006). The Development of a Formative and a Reflective Scale for the Assessment of On-line Store Usability. *Journal of Systemics, Cybernetics and Informatics*. Volume 6 - Number 5. 36-41. ISSN: 1690-4524.

Clausen, G. L. (1994). A comparative study of discomfort caused by indoor air pollution, thermal load and noise. *Indoor Air: International Journal of Indoor Air Quality and Climate* , 4:255–62.

Clausen, G., &Wyon, D. P. (2008). The Combined Effects of Many Different Indoor Environmental Factors on Acceptability and Office Work Performance. *HVAC&R Research* , 14:1, 103-113.

Diamantopoulos, A., &Winklhofer, H. (2001). Index Construction with Formative Indicators: An alternative to Scale Development. *Journal Of Marketing Research* , 269-277.

Diamantopoulos, A., Riefler, P., and Roth, K. P. (2008). Advancing Formative Measurement Models. *Journal of Business Research* (61:12), pp. 1203-1218.

Diamantopoulos, A., &Riefler, P. (2011). Using Formative Measures in International Marketing Models: A Cautionary Tale Using Consumer Animosity as an Example. *Measurement and Research Methods in International Marketing*, 11–30. doi:10.1108/s1474-7979(2011)0000022004

Dijkstra, T. 1983. "Some Comments on Maximum Likelihood and Partial Least Squares Methods". *Journal of Econometrics* (22:1/2), pp. 67-90.

Grove, S. K., Burns, N., & Gray, J. (2012). The Practice of Nursing Research: Appraisal, Synthesis, and Generation of Evidence. *Elsevier Saunders*, 382-405.

Gudergan, S. P., Ringle, C. M., Wende, S., and Will, A. 2008. Confirmatory Tetrad Analysis in PLS Path Modeling. *Journal of Business Research* (61:12), 1238-1249.

Hair, J., Hult, G., Ringle, C., &Sarstedt, M. (2014). A Primer on Partial Least Squares Structural Equation Modeling (PLS-SEM). *Sage, Thousand Oaks.*

Hair, J., Ringle, C., &Sarstedt, M. (2014). *Partial Least Squares Structural Equation Modeling: Rigorous Applications, Better Results and Higher Acceptance. Long Range Planning.*

Horie, G., Y. Sakurai, T. Narguchi, & Matsubara, N. (1985). Synthesized evaluation of noise, lighting and thermal conditions in a room. *ProceedingsofNoise Control 85* , 491–6.

INP. (2012). *Instituto Nacional de Prevención Chile, Curso ergonomia ambiental, Niveles de Confort.*

Jarvis, C. B., MacKenzie, S. B., & Podsakoff, P. M. (2003). A critical review of construct indicators and measurement model misspecification in marketing and consumer research. *Journal of Consumer Research*, 30(2), 199–218.

Li Huang, Y. Z., & Qin Ouyang, B. C. (2012). A study on the effects of thermal, luminous, and acoustic environments on indoor environmental comfort in offices. *Building and Environment* , 304-309.

MacKenzie, S. B., Podsakoff, P. M., and Jarvis, C.B. (2005). The problem of measurement model misspecification in behavioral and organizational research and some recommended solutions. *Journal of Applied Psychology* 90 (4), 710-730.

MacKenzie, S. B., Podsakoff, P. M., and Podsakoff, N. P. (2011). Construct Measurement and Validation Procedures in MIS and Behavioral Research: Integrating New and Existing Techniques,  *MIS Quarterly* (35:2), 293-334.

Olesen, B., &Seelen, J. (1993). Criteria for a Comfortable Indoor Environment in Building. *IndoorClimate* , 545-549.

OMS. (2013). *Ambientes de Trabajo Saludables: un modelo para la acción* . Suiza: Organización Mundial de la Salud.

Pérez, J., & Fidalgo, M. (1994). *NTP 394: Satisfacción laboral: escala general de satisfacción. Madrid: INSHT.*

Pett, M. A., Lackey, N. R., & Sullivan, J. J. (2003). *Making Sense of Factor Anáyisis: The Use of Factor Analysis for Instrumen Development in Health Care Research.*Sage Publications.

Rey, F., & Velasco, E. (2007). Calidad de ambientes interiores.

Ringle, Christian M., Wende, Sven, & Becker, Jan-Michael. (2015). SmartPLS 3. Bönningstedt: SmartPLS. Retrieved from http://www.smartpls.com

Rohles, F. J. (1989). Indoor environment acceptability: The development of a rating scale. *ASHRAE Transactions* , 95(1):23–27.

Sullivan, G. M., &Feinn, R. (2012). Using Effect Size—or Why the P Value Is Not Enough. *JournalofGraduate Medical Education*, 4 (3), 279–282. doi:10.4300/jgme-d-12-00156.1

UNE-EN ISO 7730:2006 "Ergonomía del ambiente térmico. Determinación analítica e interpretación del bienestar térmico mediante el cálculo de los índices PMV y PPD y los criterios de bienestar térmico local "

Warr, P., Cook, J., & Wall, T. (1979). Scales for the measurement of some work attitudes and aspects of psychological well-being. *Journal of occupational and organizational psychology* .

Wong, L., Mui, K., & Hui, P. (2008). *A multivariate-logistic model for acceptance of indoor environmental quality (IEQ) in offices.* Hong Kong, China: Elsevier Science, Building and Environment, Volume 43, Issue 1.

Woods, J. G. (1987). Office worker perceptions of indoor air quality effects on discomfort and performance. *Proceedings of Indoor Air* , 464-68.

World Bank. (2017). *The World Bank Data*. Retrieved from Urban Population 1960-2016: https://data.worldbank.org/indicator/SP.URB.TOTL